\documentclass[aps,prd,floats,floatfix,amssymb,twocolumn,superscriptaddress,nofootinbib]{revtex4-1}

\usepackage{amsmath}
\usepackage{amsfonts,amssymb}
\usepackage{graphicx,graphics}
\usepackage{hyperref}

\begin{document}

\title{Analogue models for Schwarzschild and Reissner-Nordstr\"om spacetimes}

\author{Christyan C. de Oliveira}
\email{chris@ifi.unicamp.br}
\affiliation{IFGW, Universidade Estadual de Campinas, 13083-859, Campinas, SP, Brazil}
\author{Ricardo A. Mosna}
\email{mosna@unicamp.br}
\affiliation{Departamento de Matem\'atica Aplicada, Universidade Estadual de Campinas, 13083-859, Campinas, SP, Brazil}
\author{Jo\~ao Paulo M. Pitelli}
\email{pitelli@unicamp.br}
\affiliation{Departamento de Matem\'atica Aplicada, Universidade Estadual de Campinas, 13083-859, Campinas, SP, Brazil}
\author{Maur\'icio Richartz}
\email{mauricio.richartz@ufabc.edu.br}
\affiliation{Centro de Matem\'atica, Computa\c{c}\~ao e Cogni\c{c}\~ao, UFABC, 09210-170 Santo Andr\'e, SP, Brazil}

\begin{abstract}
We present analogue models for the Schwarzschild and Reissner-Nordstr\"om (RN) spacetimes based on unidirectional hydrodynamic flows. We show that, with appropriate coordinate transformations, sound  waves in a moving fluid propagate as scalar fields at the equatorial sections of Schwarzschild and RN black holes. The coordinate associated with the direction of the flow plays the role of the radial coordinate in standard Schwarzschild/RN coordinates, while the transversal spatial coordinate is related to the angular coordinate on each equatorial slice. The magnitude of the flow velocity is related to the mass and charge of the analogue black hole. Physical quantities like pressure and density remain finite at the analogue horizon, thereby resulting in sound waves that are well defined almost everywhere, except possibly at the analogue black hole singularity.
\end{abstract}

\maketitle

\section{Introduction}
\label{sec:intro}

Black holes are among the most magnificent and extraordinary objects in physics. They constitute spacetime regions of extreme gravitational effects, from where nothing, not even light, can escape. In general relativity (GR), and in its extensions, several types of black holes (which depend on the matter fields, the dimension of the spacetime, and the symmetries assumed) have been identified and studied. For both theoretical and practical reasons, by far the most important and most explored class of black holes is the class of stationary and axisymmetric solutions of the Einstein-Maxwell equations in four dimensions. These black holes, known as Kerr-Newman (KN) black holes, are uniquely characterized by three quantities: their mass $M$, their charge $Q$, and their angular momentum $J$~\cite{Chrusciel:2012jk}. Schwarzschild ($Q=J=0$), Reissner-Nordst\"om (RN) ($J=0$), and Kerr ($Q=0$) black holes belong to the KN class.  

Even though weak field tests of GR have been performed almost since the inception of the theory~\cite{Will:2014kxa,2019NatPh..15..416C}, the strong field regime (especially in the vicinity of a black hole) remained untested for a long time. This scenario, however, began to change recently, with the gravitational wave detections by  LIGO and VIRGO~\cite{Abbott:2016blz}, and with the observation of the shadow of the black hole in the center of the galaxy M87 by the Event Horizon Telescope~\cite{Akiyama:2019cqa}. In complement to experimental tests of GR like these, some black hole phenomena can also be investigated in the context of analogue gravity~\cite{Barcelo:2005fc,Barcelo:2018ynq}. For instance, recent experiments with water, with quantum fluids of light, and with Bose-Einstein condensates have allowed the observation of Hawking radiation~\cite{Rousseaux:2007is,Weinfurtner:2010nu,Euve:2015vml,Steinhauer:2015saa,deNova:2018rld,kolobov2021observation} and of rotational superradiance~\cite{Torres:2016iee} in earth-based laboratories. Cosmological expansion~\cite{Eckel:2017uqx} and the ringdown of a black hole~\cite{Torres_2019,Torres_2020} have also been analyzed with analogue models.

Analogue Gravity, put forth by Unruh in 1981~\cite{Unruh:1980cg}, relies on the fact that sound waves on irrotational flows of inviscid fluids can mimic minimally coupled massless scalar fields around a black hole. Such fluid flows (and other physical systems which were later shown to behave in a similar fashion~\cite{Barcelo:2005fc}) are now called analogue black holes. Their corresponding line element, in a coordinate system $x^ \mu = (t,\vec{x})$, has the form  
\begin{equation}
\label{metric}
ds^ 2 =  \Omega^ 2(x^\mu)\left[-c^2(x^\mu) dt^ 2 + (d \vec{x} - \vec{v}(x^{\mu}) dt)^ 2\right],
\end{equation}
with $\vec{v}=\vec{v}(x^\mu)$ denoting the fluid velocity and $c=c(x^\mu)$ denoting the absolute value of the sound velocity in the fluid. The conformal factor $\Omega$ depends on the dimensions of the spacetime, on the density $\rho=\rho(x^\mu)$ of the fluid, and on the speed of sound. The density and the velocity are related by the continuity equation $\partial_t \rho + \nabla\cdot(\rho \vec{v}) = 0$. The analogue of the event horizon is the spatial surface whose inward unit normal $\vec{n}$ satisfies $\vec{v} \cdot \vec{n} = c$, corresponding to a surface at which the normal flow velocity is equal to the speed of sound. In fact, to obtain an analogue black hole, one has to engineer the fluid flow accordingly, so that the effective metric becomes identical to the metric of a particular black hole.  
    
    Since the coupling between the scalar field and the analogue geometry is minimal, as opposed to conformal, the factor $\Omega$ in \eqref{metric} may be important for wave-related phenomena. In the eikonal limit, however, wave propagation is well approximated by null geodesics and therefore the conformal factor $\Omega$ is irrelevant. More specifically, the occurrence of Hawking radiation and of superradiance are independent of $\Omega$~\cite{1993CQGra..10L.201J,Brito:2015oca}. In fact, the temperature of the Hawking radiation is a conformal invariant, but the greybody factors are not. Regarding superradiance, even though the frequency range that determines the occurrence of the effect is a conformal invariant, the amplification factors themselves depend on $\Omega$. 
    
Having in mind the importance of the conformal factor, together with the relevance of the KN class of black holes and the recent success of analogue gravity experiments, we ask in this paper whether or not it is possible, in principle, to construct exact (not only conformal) analogues of the Schwarzschild and RN black holes.  
Analogue black holes whose metric is conformal to the Schwarzschild and RN metrics have already been discussed in the literature in terms of Painlev\'e-Gulstrand coordinates, see e.g.~Refs.~\cite{Visser:1997ux,Barcelo:2004wz,Hossenfelder:2014gwa}. Even exact analogues of Schwarzschild black holes in isotropic coordinates have been previously obtained with static fluid flows in \cite{2005CQGra..22.2493V}. 
The procedure works for any spherically symmetric spacetime geometry and, therefore, can be applied to the RN case as well.

In this work we use a different strategy to obtain analogues, in standard coordinates, of the Schwarzschild and RN metrics. The flows considered here have \emph{one-dimensional velocity profiles that point to a fixed direction} in the laboratory frame. By judiciously
transforming and identifying the laboratory coordinates, we reproduce exactly an equatorial slice of Schwarzschild and RN metrics, with the right conformal factor. Although physical quantities like density and pressure diverge at the analogue curvature singularities, they remain finite and well behaved at the spacetime horizons.
 
\section{The model}
\label{sec:cartesian}

Let $(x,y)$ denote Cartesian coordinates on an inertial laboratory frame whose time is measured by $t$ and assume an inviscid flow whose velocity field is given by 
\begin{equation}
\vec{v}=v(x)\hat{x}.
\label{v linear}
\end{equation}
 With two spatial dimensions, the conformal factor of \eqref{metric} takes the form $\Omega = \rho(x) / c(x)$~\cite{Barcelo:2005fc}. The continuity equation, on the other hand, implies that
 \begin{equation}
 \rho(x)=\frac{k}{v(x)}, \label{rho}
\end{equation}
where $k$ is a constant. Hence, we can associate this flow with the metric
\begin{equation}
ds^2=\frac{k^ 2}{c^ 2v^ 2}\left[-(c^2-v^2)dt^2-2vdtdx+dx^2+dy^2\right].   
\label{effective density}
\end{equation}

Our first goal in this paper is to  recast the analogue metric above as the equatorial section of a typical static and spherically symmetric spacetime whose metric, in standard coordinates $(T,R,\Theta)$, is
\begin{equation}
ds^2=-f(R)dT^2+\frac{1}{f(R)}dR^2+R^2d\Theta^2.
\label{analogue metric}
\end{equation}

First, to eliminate the nondiagonal terms in \eqref{effective density}, we define a new time coordinate $T$ by 
\begin{equation}
\label{eq_for_T}
T=t+\int{\frac{v(x')}{c^2(x')-v^2(x')}dx'},
\end{equation}
which transforms the effective metric into
\begin{equation}
ds^2=\frac{k^ 2}{c^ 2v^ 2}\left[-(c^2-v^2)dT^2+\frac{c^2}{c^2-v^2}dx^2+dy^2\right].
\label{metric redefinition}
\end{equation}

Second, to reproduce the angular part of the metric (\ref{analogue metric}), we define a new spatial coordinate $R=R(x)$ according to
\begin{equation}
R(x)=\frac{k}{c(x)v(x)}.
\label{Rdev}
\end{equation}

Third, if we impose periodic boundary conditions\footnote{This limitation of the model can be overcome by assuming polar or spherical coordinates from the beginning. As shown in Appendix \ref{appendix_polar}, the procedure described here also works when the generic analogue metric \eqref{effective density} is written in polar coordinates and the background flow is assumed to be purely radial.} on $y$ so that $y\equiv y+L$, where $L$ is some fixed characteristic length, and define a new coordinate $\Theta=y/L \, (\text{mod}2\pi)$, the acoustic metric (\ref{metric redefinition}) becomes, in the $(T,R,\Theta)$ coordinates,
\begin{equation}
	ds^2 =\left(R^2 v^2 -\frac{k^2}{v^2}\right)dT^2 +\frac{dR^2}{\frac{R'^2}{R^2}-\frac{v^4R'^2}{k^2}}+R^2d\Theta^2,
	\label{metric before}
\end{equation}
where $v=v(x(R))$ and $R' \equiv dR/dx$.

Finally we require  the flow velocity to be
\begin{equation}
v=\pm k \frac{R(x)}{R'(x)}.
\label{vdeR}
\end{equation}
This guarantees that the coefficients of $dT^2$ and of $dR^2$ are the inverse of each other with flipped signs, and thus the analogue metric becomes \begin{equation}
	ds^2 \!=\!-\! \left(\!\frac{R'^2}{R^2}-\frac{k^2R^4}{R'^2}\!\right) \! dT^2 + \frac{dR^2}{\frac{R'^2}{R^2}-\frac{k^2R^4}{R'^2}}
+R^2d\Theta^2.
	\label{metric after}
\end{equation}

Therefore, any metric given by \eqref{analogue metric}, with arbitrary $f(R)$, can be matched with the analogue metric above provided that the coordinate change $R=R(x)$ satisfies the nonlinear ordinary differential equation
\begin{equation}
R'^4-f(R) R^2 R'^2-k^2 R^6=0.
\label{edo geral}
\end{equation}
Since this equation is invariant under a translation of the laboratory coordinate, i.e.~$x\to x+\text{const}$, we have considerable freedom to choose the associated boundary condition.

We are particularly interested in mimicking the equatorial section of the Schwarzschild and RN metrics in (1+2) fluid analogues. To accomplish that, in the next sections we will solve \eqref{edo geral} using the corresponding function $f(R)$ for each metric of interest. This determines the coordinate transformation $R=R(x)$, from which the flow and sound velocities can be calculated by means of Eqs.~(\ref{Rdev}) and  (\ref{vdeR}). Once the velocities are known, the pressure $P=P(x)$ can be found by integrating
\begin{equation}
\frac{dP}{dx}=c^{2}(x) \frac{d\rho}{dx}.
\label{state eq}
\end{equation}
If desired, the function $F$ that defines the equation of state $F(P,\rho)=0$ can then be obtained by eliminating the dependence on $x$ from $P(x)$ and $\rho(x)$ simultaneously. In particular, if the coordinate $x$ is restricted to a domain where $\rho(x)$ is one-to-one, an equation of state in the form $P=P(\rho)$ is found.
Additionally, the external force $\vec{f}$ required to maintain the desired flow is given by 
\begin{equation}
\vec{f}=\rho\frac{d\vec{v}}{dt}+\nabla P =  k\frac{dv}{dx}\left(1-\frac{c^ 2}{v^ 2} \right)\hat{x},
\end{equation}
where the first equality is the Euler's equation and we have used Eqs.~\eqref{rho} and \eqref{state eq} to derive the second equality.  

\section{Equatorial section of Schwarzschild}

The equatorial section of a Schwarzschild spacetime of mass $M$, in standard coordinates, is given by \eqref{analogue metric} with $f(R)=1-R_{S}/R$,
where $R_{S}=2GM/c_\ell^2$ is the Schwarzschild radius, $G$ the gravitational constant, and $c_\ell$ is the speed of light. To simplify the notation, we adopt units in which $R_S=1=k^2$. Equation~(\ref{edo geral}) then becomes
\begin{equation}
R'^4 +  (R  -  R^{2}) R'^ 2  - R^{6}  = 0,
\label{edo sch cart}
\end{equation}
which we solve numerically assuming the boundary condition $R(0)=1$. This boundary condition fixes the Schwarzschild horizon to be at $x=0$ in the laboratory coordinates. 

\begin{figure}[th!]
\centering
\includegraphics{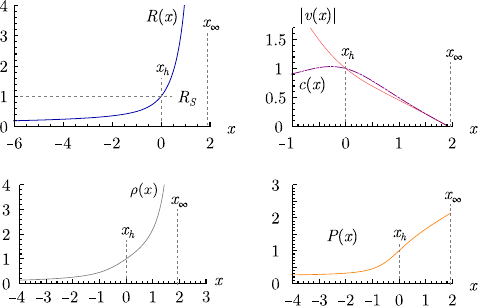}
\caption{Analogue model for the equatorial section of a Schwazschild black hole. Top left: numerical solution $R(x)$ of Eq.~(\ref{edo sch cart}). The horizon at $R=R_S=1$ corresponds to $x=x_h=0$ in the laboratory frame. The external region of the black hole is covered by the finite interval $(x_h,x_\infty)$ in the laboratory coordinate $x$. Top right: flow velocity $v(x)$ and sound velocity $c(x)$; the location where $|v(x)|=c(x)$ corresponds to the Schwarzschild horizon. Bottom: density $\rho(x)$ and pressure $P(x)$ associated with the numerical solution.}
\label{fig:cartesian schwarzschild 2d}
\end{figure}

The numerical solution that increases monotonically with $x$ is shown in Fig.~\ref{fig:cartesian schwarzschild 2d}(top left). Following the discussion in the previous section, the corresponding fluid velocity, sound velocity, density of the fluid and pressure in the flow, respectively $v(x)$, $c(x)$, $\rho(x)$ and $P(x)$, can be straightforwardly determined from the numerical solution $R(x)$. The results are also shown in Fig.~\ref{fig:cartesian schwarzschild 2d}.
Another solution for $R(x)$, which decreases with $x$, also exists, but there is no significant difference when compared to the situation analyzed here.

All horizontal axes in Fig.~\ref{fig:cartesian schwarzschild 2d} represent the linear spatial coordinate $x$ at the laboratory. The singularity of the Schwarzschild metric at $R=0$  corresponds to $x\to-\infty$ in the laboratory frame. Spatial infinity of the Schwarzschild spacetime, i.e.~$R\to+\infty$, corresponds to a finite value of the laboratory coordinate denoted by $x_{\infty}$. Moreover, the Schwarzschild event horizon, located at $x=0$, is also (by construction) the point at which the velocities of the flow and of the waves have the same magnitude.

 The choice of sign in Eq.~(\ref{vdeR}) determines the direction of the flow and, therefore, whether the event horizon corresponds to a black hole or to a white hole. Note that the entire external region outside the horizon is mapped into a finite range of the laboratory coordinates. We highlight that all relevant quantities, shown in Fig.~\ref{fig:cartesian schwarzschild 2d}, are well behaved everywhere, including at the horizon. 

\section{Equatorial section of Reissner-Nordstr\"om}
The metric of the equatorial section of a RN spacetime of mass $M$ and electric charge $Q$, in standard coordinates, is also \eqref{analogue metric}, but now $f(R)$ is given by $f(R) = 1-R_{S}/R + R_{Q}^{2}/R^{2}$. Here $R_S$ is the same as in the Schwarzschild case and $R_Q=G^{1/2} Q/c_\ell^2$ is a characteristic length fixed by the charge. In terms of the mass-to-charge ratio $m=R_S/R_Q$, the existence and locations of the horizons are determined by the roots
\begin{equation}
R_{\pm} = \frac{1}{2}\left( m \pm \sqrt{m^{2} - 4} \right)
\end{equation}
of $f(R)$, where we chose units in which $R_Q=1=k^ 2$. There are three cases to consider:
\begin{enumerate}
	\item[(i)]   $ m > 2 $, for which there are two horizons (matter dominates, $R_{S}>2R_{Q}$);
	
	\item[(ii)]  $ m = 2 $, for which there is only one horizon (extreme case, $R_{S}=2R_{Q}$);
	
	\item[(iii)] $ m < 2 $, for which there is no horizon and the RN spacetime has a naked singularity at $R=0$ (charge dominates, $R_{S}<2R_{Q}$).
\end{enumerate}

\begin{figure}[th!]
\centering
\includegraphics{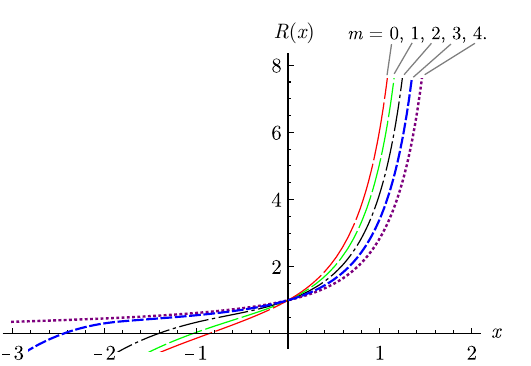}
\caption{Analogue model for the equatorial section of a RN spacetime. The plot shows the numerical solutions for $R(x)$ from Eq.~(\ref{cartesian rn equation}) for several values of $m=R_S/R_Q$.}
\label{fig: rn various ms}
\end{figure}

The differential equation \eqref{edo geral} that yields the coordinate transformation $R=R(x)$ is now given by
\begin{equation}
R'^{4}(x)  + \left[ mR(x) - R^{2}(x) - 1 \right] R'^{2}(x) - R^{6}(x) = 0,
\label{cartesian rn equation}
\end{equation} 
which we solve numerically for the boundary condition $R(0)=1$. The results for several values of $m$ are exhibited in Fig.~\ref{fig: rn various ms}.
Similarly to the Schwarzschild case, for each value of $m$ the function $R(x)$ approaches infinity at a finite value $x_{\infty}$ of the laboratory coordinate. However, differently from the Schwarzschild case, these solutions now become negative if $x$ is less than a certain $x_0$ which depends on $m$. Since we are interested in interpreting $R(x)$ as the radial coordinate of the RN metric, we restrict ourselves to $x>x_0$. As a result, the RN spacetime is mapped to a finite interval of the laboratory coordinate, given by $x_0 < x < x_{\infty}$.

\begin{figure}[h]
\centering
\includegraphics{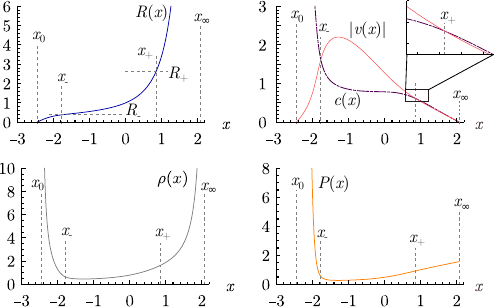}
\caption{Analogue model for the equatorial section of a RN black hole. Top left: numerical solution of $R(x)$ from Eq.~(\ref{cartesian rn equation}), for $m=3$. The whole range of the radial coordinate $R$ is covered by a finite interval $(x_0,x_\infty)$ in the laboratory frame. Top right: flow velocity $v(x)$ and sound velocity $c(x)$; the locations of the two horizons correspond to the points $x_-$ and $x_+$ wherein $|v(x)|=c(x)$. Bottom: density $\rho(x)$ and pressure $P(x)$ associated with the numerical solution.}
\label{fig: rn m=3}
\end{figure}

\begin{figure}[h]
\centering
\includegraphics{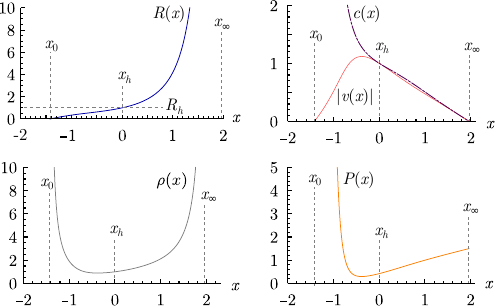}
\caption{Analogue model for the equatorial section of an extremal RN black hole. Top left: numerical solution of $R(x)$ from Eq.~(\ref{cartesian rn equation}), for $m=2$. The whole range of the radial coordinate $R$ is covered by a finite interval $(x_0,x_\infty)$ in the laboratory frame. Top right: flow velocity $v(x)$ and sound velocity $c(x)$; the location of the horizon corresponds to the point $x_h$ wherein $|v(x)|=c(x)$. Bottom: density $\rho(x)$ and pressure $P(x)$ associated with the numerical solution.}
\label{fig: rn m=2}
\end{figure}

\begin{figure}[h!]
\centering
\includegraphics{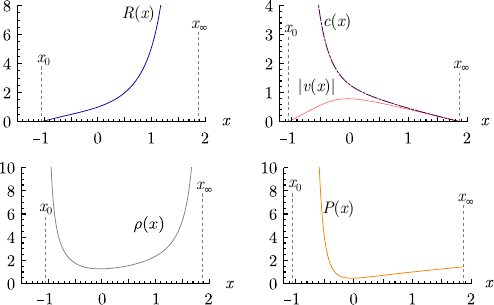}
\caption{Analogue model for the equatorial section of a RN naked singularity. Top left: numerical solution of $R(x)$ from Eq.~(\ref{cartesian rn equation}), for $m=1$. The whole range of the radial coordinate $R$ is covered by a finite interval $(x_0,x_\infty)$ in the laboratory frame. Top right: flow velocity $v(x)$ and sound velocity $c(x)$.  Bottom: density $\rho(x)$ and pressure $P(x)$ associated with the numerical solution.}
\label{fig: rn m=1}
\end{figure}

Once the solution $R(x)$ is obtained, the other parameters describing the flow are easily determined as previously explained. We apply this procedure to a nonextremal black hole ($m=3$), to an extremal black hole ($m=2$), and to a naked singularity ($m=1$). The results are shown in Figs.~\ref{fig: rn m=3},  \ref{fig: rn m=2}, and \ref{fig: rn m=1}, respectively. Once again the horizons of the spacetime correspond, in the laboratory, to the locations where (the absolute value of) the fluid velocity equals the sound velocity. We see that there are two such locations in the top right panel of Fig.~\ref{fig: rn m=3}, one in Fig.~\ref{fig: rn m=2} and none in Fig.~\ref{fig: rn m=1}, as expected from the discussion in the beginning of this section.

 Moreover, in the case of Fig.~\ref{fig: rn m=3}, we see that $|v(x)|>c(x)$ for $x_-<x<x_+$, which corresponds to the region between the two horizons, $R_{-}$ and $R_{+}$.

 As expected, the density and pressure diverge at the location $x_0$ corresponding to the singularity of the black hole. Nevertheless, all the physical quantities associated with the flow remain perfectly well behaved in the neighborhood of the horizons, which is an appealing feature of the model.

\section{Final remarks}

 We have proposed and analyzed a procedure to find two-dimensional fluid flows that mimic the equatorial section of spherically symmetric spacetimes, not only conformally but exactly. In particular, the equator of a Schwarzschild spacetime and of a RN spacetime have been obtained in the context of analogue gravity for unidirectional fluid flows. Regarding the RN class of spacetimes, not only black holes (both extremal and nonextremal), but also naked singularities, can be reproduced. Physical quantities, namely density, pressure, flow velocity, and sound velocity, have been obtained numerically and have been shown to be well behaved everywhere, except at the laboratory coordinate that corresponds to the singularity of the simulated spacetime. In Appendix \ref{appendix_polar}, a similar procedure, with similar results, is discussed for radial flows in two and in three spatial dimensions. Three-dimensional flows, in particular, can mimic not only the equator of the Schwarzschild and RN spacetimes, but their full 1+3 metrics. 
 
Concerning the possibility of experimental realization of the analogue metrics discussed in this paper, even though a detailed analysis is beyond the scope of our work, we would like to point out that the method developed here may also be applied, with appropriate modifications, to gravity waves propagating on the surface of a shallow basin with a curved bottom and filled with a liquid~\cite{schutzhold2002gravity}. In that case, the task of providing a variable density $\rho(x)$ and a variable pressure $P(x)$ satisfying a specific equation of state is replaced by the task of providing a variable fluid depth $h(x)$ and a specific nonflat bottom for the basin. This could provide an alternative framework, perhaps closer to experimental realization, wherein our most appealing result, which is the possibility of an exact match with the equatorial section of a spherically symmetric metric, still holds.

To conclude, we remark that in order to go beyond the results of Refs.~\cite{Visser:1997ux,Barcelo:2004wz,Hossenfelder:2014gwa}, in which analogue metrics that are conformal to either the Schwarzschild or the RN metric were obtained, one has to explore the freedom in the choice of the radial coordinate. This extra degree of freedom allows one to determine exact analogues of spherically symmetric spacetimes, as we have done in this work. We point out that such exact analogues, in isotropic coordinates, have been previously obtained in Ref.~\cite{2005CQGra..22.2493V} for static flows. Even though their approach is different from ours, one similarity is the fact that the exact match with the spacetime metric requires some fine-tuning (in particular, of the equation of state and of the external force needed to maintain the background flow). Interestingly enough, these two models (that of Ref.~\cite{2005CQGra..22.2493V} and the one presented here) are essentially unique, in a sense which is made more precise in Appendix \ref{appendix_uniqueness}.

\acknowledgments
C. C. O. acknowledges support from the Conselho Nacional de Desenvolvimento Cient\'{i}fico e Tecnol\'{o}gico (CNPq, Brazil), Grant No. 142529/2018-4. R. A. M. was partially supported by Conselho Nacional de Desenvolvimento Científico e Tecnológico under Grant No. 310403/2019-7. M.~R.~acknowledges support from the Conselho Nacional de Desenvolvimento Cient\'{i}fico e Tecnol\'{o}gico (CNPq, Brazil), Grants No. FA 309749/2017-4 and No. FA 315664/2020-7.

\appendix

\section{Polar and spherical coordinates}
\label{appendix_polar}

The model presented in the main text requires the identification of the Cartesian coordinate $y$ with the angular coordinate $\Theta$ of the equatorial section of the spherically symmetric spacetime under investigation. Here we show that this can be avoided by working directly with polar $(r,\theta)$ or spherical $(r,\theta,\phi)$ coordinates in the laboratory frame, and by considering, instead of (\ref{v linear}), a radial profile
\begin{equation}
\label{v radial}
\vec{v}=v(r) \, \hat{r}
\end{equation}
for the velocity field. As we show next, the angular coordinate of the spacetime can then be identified with a \emph{bona fide} angular coordinate of the laboratory. In particular, if the analogue model is (1+3)-dimensional and we employ spherical coordinates, the full spherically symmetric metric, and not only its equatorial section, can be recovered.

\subsection{Polar coordinates in 2D}
\label{sec:polar}
For a two-dimensional flow with a purely radial velocity profile given by \eqref{v radial}, the continuity equation yields
\begin{equation}
\rho(r) = \frac{k}{r v(r)},
\label{polar density}
\end{equation}
where $k$ is a constant. This leads to an acoustic metric of the form
\begin{equation}
ds^{2} = \frac{k^{2}}{r^{2}v^{2}c^{2}} \left[  -(c^{2} -v^{2} ) dT ^{2} + \frac{c^{2}}{c^{2} - v^{2}} dr^{2} + r^{2} d\theta^{2} \right]
\label{polar line element}
\end{equation}
in terms of polar coordinates. Once again, the laboratory time $t$ was redefined to $T$ by
\begin{equation}
T = t + \int \frac{v(r')dr'}{c^{2}(r') -v^{2} (r')}.
\end{equation}

Repeating the idea used in the main text, we define a new radial coordinate $R(r)$ according to
\begin{equation}
R(r) = \frac{k}{c(r) v(r)},
\label{R(r)}
\end{equation} 
but we now take (the angular coordinate in the analogue model as) $\Theta=\theta$. Hence the analogue metric is transformed into
\begin{align}
	ds^{2} &= \frac{R^{2} v^{4}(R) - k^{2} }{r^{2} v^{2}(R)} dT^{2} \nonumber \\
	&  +  \frac{k^{2} R^{2}}{ r^2 \left( k^{2} -  R^{2} v^{4}(R) \right) R^{'2} } dR^{2} + R^{2} d\Theta^{2}. 
	\label{polar line before}
\end{align}

\begin{figure}[th!]
	\centering
\includegraphics[width=0.95\columnwidth]{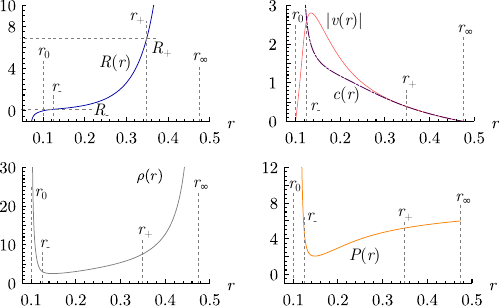}
	\caption{(2+1)-dimensional analogue model (with radial flow) of an equatorial section of the RN  spacetime. Top left: numerical solution of Eq.~(\ref{edo polar}) for a charge-to-mass ratio $m=7$. The boundary condition used is $R(0.1)=0$. Top right: radial velocity $v(r)$ and sound velocity $c(r)$; the locations where $|v(r)|=c(r)$, denoted by $r_-$ and $r_+$, correspond to the horizons. Bottom: density $\rho(r)$ and pressure $P(r)$ associated with the numerical solution.}
	\label{fig:polar2D}
\end{figure}

If we further require that the coefficients of $dT^2$ and $dR^2$ be the inverse of each other with flipped signs, we need to guarantee that the flow velocity satisfies
\begin{equation}
v(r) = \pm \frac{k R(r)}{r^{2} \, R'(r)}.
\label{polar velocity}
\end{equation}
After imposing the restriction above, the analogue metric (\ref{polar line before}) becomes exactly (\ref{analogue metric}) provided that
\begin{equation}
\label{edo polar}
r^{8} R'^{4} - f(R) r^6 R^2 R'^2 - k^2 R^6=0.
\end{equation}

This nonlinear differential equation can be solved numerically and its solution can be used to find the relevant physical quantities associated with the flow, as we did in the main text. For the sake of illustration we apply the method for a RN spacetime with charge-to-mass ratio $m=7$, which is one of the cases discussed in Sec.~V. Units are chosen such that $k^2=1$ and $f(R) = 1-\frac{m}{R}+\frac{1}{R^2}$. To solve Eq.~\eqref{edo polar} numerically, we impose the boundary condition $R(0.1)=0$, thus fixing the singularity of the spacetime to be at $r=r_0=0.1$. As in the Cartesian analysis, the range of the spacetime radial coordinate $R$ is mapped to a finite interval of the laboratory coordinate (in this case, $r$).
 The results are exhibited in Fig.~\ref{fig:polar2D}. Notice that, once again, all calculated quantities are well behaved at the horizons.

\subsection{Spherical coordinates in 3D}	
\label{sec:spherical}

For a three-dimensional flow with a purely radial velocity profile given by Eq.~(\ref{v radial}), the analogue metric (\ref{metric}) is given, in spherical coordinates, by
\begin{align}
\label{spherical metric before}
	ds^{2}& =\left(\frac{k}{r^{2}vc}\right) \left[ -(c^{2} -v^{2} ) dT ^{2} + \right.  \nonumber \\ 
	& \left.   \frac{c^{2}}{c^{2} - v^{2}} dr^{2} + r^{2}d\theta ^{2} + r^{2} \sin^{2} \theta d\phi ^{2} \right],
\end{align}  
where we have used the fact that the conformal factor is given by $\Omega = \sqrt{\rho/c}$~\cite{Barcelo:2005fc} (the continuity equation has also been used to eliminate $\rho(r)$ in terms of $v(r)$) and, again, we redefined the laboratory time $t$ as 
\begin{equation}
	T=t + \int \frac{v(r')dr'}{c^{2}(r')-v^{2}(r')}
\end{equation}
to eliminate cross terms in the analogue metric.

Following the ideas used in the main text, we see that the appropriate redefinition of the laboratory radial coordinate $r$ is now given by 
\begin{equation}
R(r) = \sqrt{\frac{k}{c(r)v(r)}},
\label{spherical R}
\end{equation}
where $k$ is once again a constant.

Finally, if we impose the additional restriction
\begin{equation}
v(r)= \pm \frac{k}{r^{2} R'(r)},
\label{spherical v}
\end{equation}
the analogue metric (\ref{spherical metric before}) transforms into
\begin{equation}
ds^{2} = - f(R) dT^{2} + \frac{1}{f(R)} dR^{2} + R^{2} d \Theta ^{2} + R^{2} sen^{2} \Theta \, d\Phi ^{2}
\end{equation}
provided that $R(r)$ is a solution of the following differential equation:
\begin{equation}
k^{2}R^{4} + f(R) r^{6} R^{2} R'^{2} - r^{8} R'^{4}=0,
\label{spherical f}
\end{equation}
and the angular coordinates are kept unchanged, i.e.~$\Theta=\theta$ and $\Phi=\phi$.

\begin{figure}[th!]
	\centering
\includegraphics[width=0.95\columnwidth]{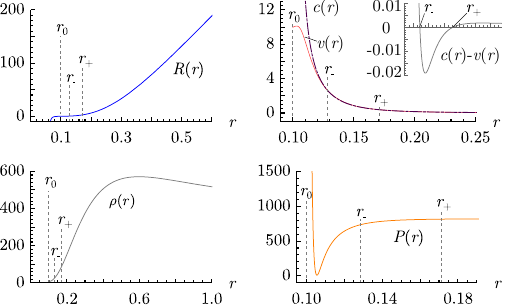}
	\caption{(3+1)-dimensional analogue model (with radial flow) of a RN spacetime.  Top left: numerical solution of Eq.~(\ref{spherical f}) for a charge-to-mass ratio $m=3$. The boundary condition used is $R(0.1)=0$. Top right: radial velocity $v(r)$ and sound velocity $c(r)$; the locations where $|v(r)|=c(r)$, denoted by $r_-$ and $r_+$, correspond to the horizons. The difference $c(r)-v(r)$ in the region $r_{-}\le r \le r_{+}$ is also plotted. Bottom: density $\rho(r)$ and pressure $P(r)$ associated with the numerical solution.}
	\label{fig:spherical3D}
\end{figure} 

By solving Eq.~(\ref{spherical f}) it is possible to reproduce the Schwarzschild and RN spacetimes exactly. As an example, we study one more time a RN black hole with charge-to-mass ratio $m=3$. Units are chosen such that $k^2=1$ and $f(R) = 1-\frac{m}{R}+\frac{1}{R^2}$. The boundary condition used to solve Eq.~\eqref{spherical f} is $R(0.1)=0$. The corresponding results are shown in Fig.~\ref{fig:spherical3D}, where we see again that all calculated quantities are well behaved at the horizons.

\section{Uniqueness of the model}
\label{appendix_uniqueness}

In this Appendix we discuss the uniqueness of the model presented here. Our aim is to find all the analogue models of the kind discussed in this paper that mimic the metric \eqref{analogue metric}. The generic analogue metric in 1+2 dimensions is given by
\begin{equation}
ds^2=\frac{\rho^ 2}{c^ 2}\left[-(c^2-v^2)dt^2-2vdtdx+dx^2+dy^2\right], 
\label{analogue_general}
\end{equation}
where $(t,x,y)$ are Cartesian coordinates in the laboratory. Our assumptions are that (1) $v$, $c$ and $\rho$ are functions of $x$ only and (2) $\Theta = y \, (\mathrm{mod} L)$, where $L$ is some fixed characteristic length, as before.

Since $\Theta = y \, (\mathrm{mod} L)$, we have $d\Theta = dy$. Comparing \eqref{analogue metric} to \eqref{analogue_general} we conclude that $R=\rho/c$ must be a function of $x$ only. Therefore, $dR = R'(x)dx$. It is evident that $T$ cannot depend on $y$ and thus $dT=T_t dt + T_x dx$. Substituting $dT$, $dR$ and $d\Theta$ in terms of $dt$, $dx$, $dy$  in \eqref{analogue metric} yields
\begin{multline}
ds^2=-f(R)T_t^2 dt^ 2 - 2f(R)T_t T_x dt dx +\\ 
+ \left(-f(R)T_x ^ 2 +\frac{R'{}^2}{f(R)}\right)dx^2 + R^2 dy^ 2.
\label{eq_ap1}
\end{multline}
It follows from Eq.~\eqref{analogue_general}  that
\begin{align}
R =& \frac{\rho}{c}, \label{apeq1} \\
-f(R)T_t^2 =& - \frac{\rho^ 2(c^2-v^2)}{c^ 2}, \label{apeq2} \\
- 2f(R)T_t T_x =& - 2 v \frac{\rho^ 2}{c^ 2}, \label{apeq3} \\
\left(-f(R)T_x ^ 2 +\frac{R'{}^2}{f(R)}\right) =& \frac{\rho^ 2}{c^ 2}. \label{apeq4}
\end{align}
Equation~(\ref{apeq2}) implies that $T_t$ is a function of $x$ only. This, together with \eqref{apeq3}, implies that $T_x$ is a function of $x$ only. Therefore, $T_{xt}=T_{tx}=0$. Since $T_{tx}=0$, we conclude that $T_t$ is constant and thus
\begin{equation}
\label{apeqTt}
T_t ^ 2 = \frac{\rho^ 2(c^2-v^2)}{f(R) c^ 2} = K^ 2.
\end{equation}
This implies that $T=K t + g(x)$, with $g(x)$ a function to be determined and $K$ a constant that, without loss of generality, we take to be $1$. It follows from Eq.~(\ref{apeq3}) that
\begin{equation} 
\label{apeqgp}
g'(x)=\frac{v \rho^ 2}{c^ 2f(R)},
\end{equation}
which leads to Eq.~(\ref{eq_for_T}) for $T$ after using Eq. \eqref{apeqTt}.

Finally, it follows from Eq.~(\ref{apeq4}), after using Eqs.~\eqref{apeqgp} and \eqref{apeq1}, that
\begin{equation}
\label{apeqRp}
R'^ 2 - v^2R^ 4 -  f(R) R^ 2 =0.
\end{equation}
We are thus left with the following independent equations: \eqref{apeq1}, \eqref{apeqTt}, \eqref{apeqgp} and \eqref{apeqRp} and five unknowns: $v(x)$, $c(x)$, $\rho(x)$, $g(x)$ and $R(x)$. The continuity equation provides the fifth relation (between $v$ and $\rho$). 

If we assume $v \neq 0$, then the only possible solution is the one presented in this paper. If $v=0$ and the 3+1 dimensional analogue metric in spherical coordinates is used instead of \eqref{analogue_general}, then our procedure is compatible with the solution obtained in \cite{2005CQGra..22.2493V}.

\bibliographystyle{apsrev4-1}


%

\end{document}